\pdfoutput=1
\documentclass[twocolumn,epjc3]{svjour3}
\usepackage{graphicx}
\usepackage{xcolor}
\usepackage{SIunits}
\usepackage{multirow}
\usepackage{setspace}
\usepackage[colorinlistoftodos]{todonotes}
\newcommand{\Nuc}[2]{\ensuremath{^{#2}\mbox{#1}}}

\usepackage{lineno}
\usepackage{hyperref}
\journalname{Eur. Phys. J. C}

\begin{document}

\title{Polyethylene naphthalate film as a wavelength shifter in liquid argon detectors}

\author{M.~Ku\'zniak\thanksref{e1,addr1,addr2}
        \and
        B.~Broerman\thanksref{addr2} 
        \and
        T.~Pollmann\thanksref{addr3}
        \and
        G.~R.~Araujo\thanksref{addr3}
}
\thankstext{e1}{e-mail: mkuzniak@physics.carleton.ca}
\institute{Department of Physics, Carleton University, Ottawa, Ontario, K1S 5B6, Canada \label{addr1}
           \and
           Department of Physics, Engineering Physics, and Astronomy, Queen's University, Kingston, Ontario, K7L 3N6, Canada \label{addr2}
           \and
           Department of Physics, Technische Universit\"at M\"unchen, 80333 Munich, Germany \label{addr3}
           }
\date{Received: date / Accepted: date}
\maketitle
\begin{abstract}
Liquid argon-based scintillation detectors are important for dark matter searches and neutrino physics. Argon scintillation light is in the vacuum ultraviolet region, making it hard to be detected by conventional means. Polyethylene naphthalate (PEN), an optically transparent thermoplastic polyester commercially available as large area sheets or rolls, is proposed as an alternative wavelength shifter to the commonly-used tetraphenyl butadiene (TPB). By combining the existing literature data and spectrometer measurements relative to TPB, we conclude that the fluorescence yield and timing of both materials may be very close. The evidence collected suggests that PEN is a suitable replacement for TPB in liquid argon neutrino detectors, and is also a promising candidate for dark matter detectors. Advantages of PEN are discussed in the context of scaling-up existing technologies to the next generation of very large ktonne-scale detectors. Its simplicity has a potential to facilitate such scale-ups, revolutionizing the field.
\end{abstract}




\section{Introduction}
Ongoing development of the next generation multi-~hundred tonne and multi-ktonne liquid argon (LAr) detectors 
calls for optimizing the technology towards high light collection efficiencies. This is highly motivated especially in dark matter searches where light yield directly effects the sensitivity and background rejection capabilities. The proposed 300 tonne Global Argon Dark Matter Collaboration detector (GADMC)~\cite{gadmc}, based on a scaled-up single-~\cite{deap} or dual-phase~\cite{ds20k} approach, is capable of reaching the ``neutrino floor", i.e.~the ultimate sensitivity at which the interactions from coherent elastic neutrino-nucleus scattering become the limiting background. GADMC will probe the remaining accessible parameter space allowed for heavy WIMPs. Furthermore, LAr neutrino detectors can benefit from the additional information provided by scintillation light~\cite{Sorel,Szelc}. DUNE is a future long baseline neutrino experiment featuring a 40~ktonne LAr far detector~\cite{dune}. Although not part of the current design, a potentially lower threshold and higher light yield would improve the overall sensitivity to neutrino oscillation physics and would enhance the precision of supernova neutrino energy reconstruction.

Due to the lack of efficient photosensors capable of directly registering the vacuum ultraviolet (VUV) light  from LAr scintillation peaked at 128~nm~\cite{Heindl}, wavelength shifter (WLS) coatings, covering inner surfaces of the cryogenic volume, are used to convert VUV photons to visible. 1,1,4,4-tetraphenyl-1,3-butadiene (TPB) has been successfully used for that purpose for the past few decades~\cite{tpb1,tpb2,Benson}. TPB can be evaporatively coated, which requires a dedicated process control system and high vacuum conditions~\cite{tpbdeap}, or can be dissolved in a polymeric solution and applied as paint, which leads to a degradation in conversion efficiency by a factor $>$2--3~\cite{McKinsey,Mavrokoridis} due to VUV absorption in the polymer.

These large-scale endeavors call for a simple, inexpensive, and robust WLS solution easily scalable to surfaces ranging in size between a few hundred and a few thousand m$^2$. We show that poly(ethylene 2,6-naphthalate) (PEN) is well suited for this purpose.
\section{Polyethylene naphthalate}
PEN is a thermoplastic polyester with physical properties similar to common poly(ethylene terephthalate) or PET (mylar). It can be easily drawn into foils or extrusion moulded, and has wide applications in industry.

\begin{figure}[ht]
\centering\includegraphics[width=\linewidth]{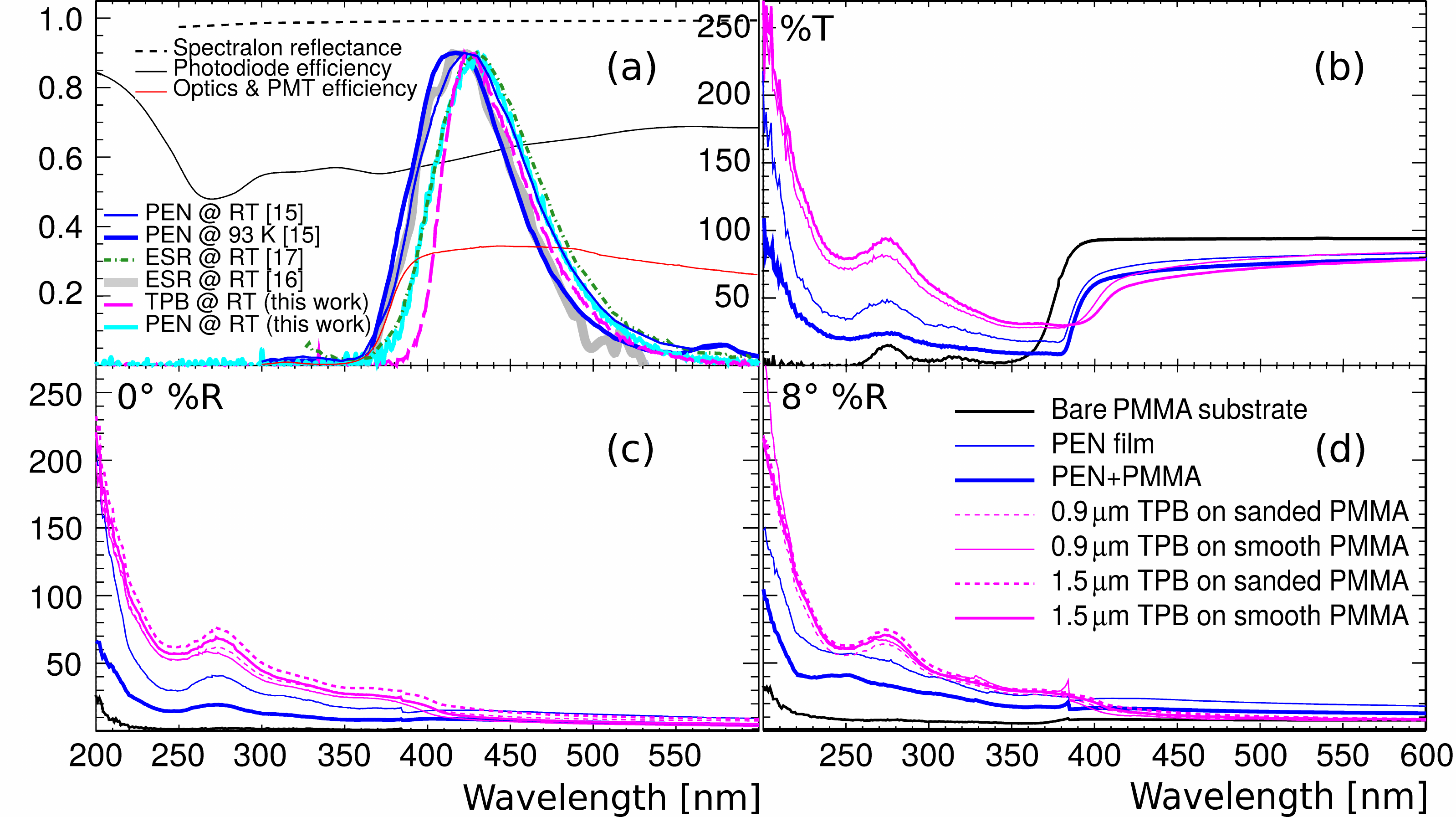}
\caption{(a) Emission spectra of PEN~\cite{Laurent}, ESR~\cite{Baudis,Janecek} and TPB normalized to the same amplitude. Also shown are the Spectralon hemispherical reflectance, integrating sphere photodiode efficiency, and combined efficiency of optics and PMT in the VUV monochromator setup (see legend). Raw spectra of (b) transmittance (T-mode), (c) diffuse reflectance (R-mode with 0$^\circ$ sample holder) and (d) hemispherical reflectance (R-mode with 8$^\circ$ sample holder) of TPB, PEN and acrylic samples (see legend). 
Acrylic blocks transmission below 250~nm and is highly transparent above 400~nm.
}
\label{fig:spectra}
\end{figure}

The main component of PEN emission is at 430~nm (Fig.~\ref{fig:spectra}(a)) and has been thoroughly investigated in the literature. It originates from excimer formation between naphthalene-dicarboxylate units which results in $^1$($\pi$,$\pi$*) fluorescence transition. Studies of fluorescence in the 50--650~nm excitation range~\cite{Ouchi06} and the temperature dependence of the fluorescence spectrum and intensity in the 93--293~K range~\cite{Laurent} have been published. 

PEN has been identified as a promising scintillator because its light yield exceeds that of conventional, expensive organic scintillators~\cite{Nakamura}. As such, it has been proposed as a radiopure self-vetoing structural material for use in ultra-low background experiments~\cite{KamlandZen,Majorovits}.

PEN together with acrylic (poly(methyl methacrylate) or PMMA) are two components of a polymeric reflector film Vikuiti\texttrademark\ Enhanced Specular Reflector (ESR, formerly known as VM2000) by 3M\textsuperscript{\textregistered}~\cite[Fig. 4 therein]{3m}, which is commonly used in ultra-low background experiments or noble liquid scintillation detectors as a specular reflector~\cite{miniclean} or substrate for TPB coatings~\cite{ardm,gerda,lariat}. Additionally, because of substantial scintillation light output even at cryogenic temperatures (15~mK), in the CRESST experiment it is used as an active scintillating veto against $\alpha$ particles~\cite{cresst}. 

ESR fluorescence with features identical to PEN and intensity $\sim$11 times lower than TPB has been reported~\cite{Baudis}. As ESR film is an optical stack with hundreds of interleaved, less than 120-nm thick, layers of PMMA and PEN, its overall conversion efficiency is suppressed by VUV loss in PMMA. The fluorescence lifetime is 20~ns~\cite{Janecek} at room temperature (RT) and at most a few times that in LAr~\cite{Baudis}; a value much smaller than the triplet lifetime of LAr, 1.3~\micro s, is needed for pulse-shape discrimination used in LAr detectors.

\section{Comparison with TPB}
TPB is currently the standard WLS used in the LAr detectors
~\cite{deap,miniclean,ardm,gerda,lariat,ds-50,microboone}.
In what follows, measurements of PEN fluorescence at RT with TPB as a reference are described. 
The integrating sphere (IS) method was chosen as less sensitive to differences in angular distributions of fluorescence from test and reference samples than a fixed angle measurement, to determine the relative yield of PEN to TPB in the 200--260~nm range and its systematic uncertainty. Literature data and, for comparison, our own fixed angle VUV measurement of the relative PEN yield wavelength dependence are then used to project the IS results to 128~nm excitation. Wavelength shifting efficiency (WLSE) of PEN for 128~nm excitation at 87~K are then inferred based on the literature, so that a comparison to TPB relevant for cryogenic conditions in the final application can be made.

\begin{figure}[ht]
  \centering
  \includegraphics[trim={0 4.2cm 0 3.7cm},clip,width=\linewidth]{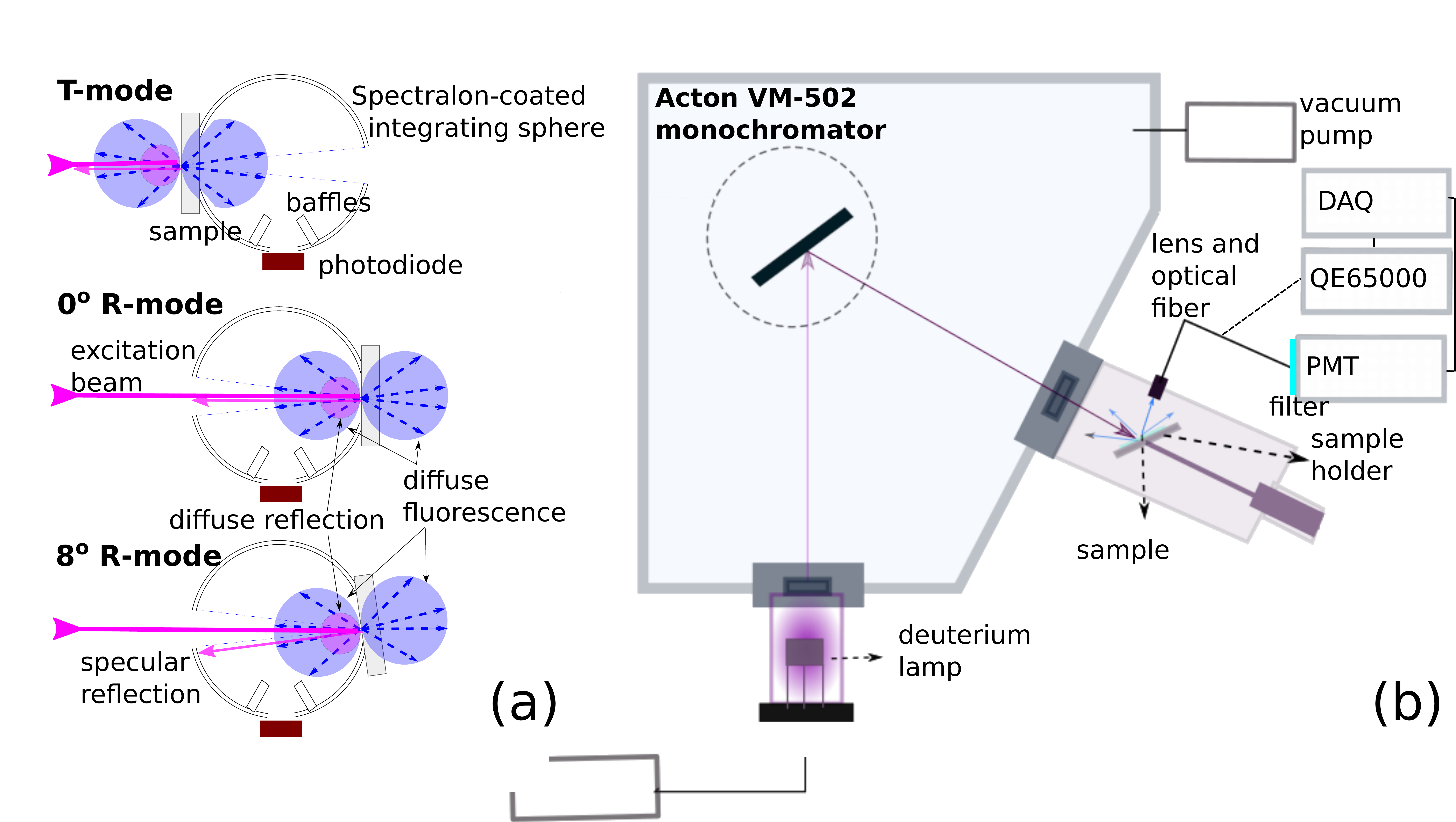} 
\caption{(a) IS configurations. T-mode: excitation wavelength is fully blocked by the substrate, only fluorescence emitted forward is registered. 0$^\circ$ R-mode: specular reflection of the incident beam escapes the IS, the diffuse reflection and fluorescence emitted back are registered. 8$^\circ$ R-mode: specular and diffuse reflection of the incident beam and the fluorescence emitted back are registered. (b) Schematic of the fixed angle VUV measurement setup. Its photon detector is either a PMT or a spectrometer.}
\label{fig:schematic}
\end{figure}

Samples of 125~\micro\meter\ thick PEN film (Teonex\textsuperscript{\textregistered}~Q83) were supplied by Teijin DuPont Films. For the IS measurement 0.9 and 1.5~\micro\metre\ thick TPB coatings were vacuum evaporated on 1/8-inch-thick UV absorbing acrylic substrates with varying surface finish (smooth or sanded), using the evaporation system described in~\cite{Tina}. To ensure a pristine state of TPB coatings, samples were stored in dark and dry conditions and the measurements were carried out within one week from their production.
Additionally, for direct comparison the PEN sample was coupled with a transparent optical grease (Saint-Gobain BC-630) to an identical smooth substrate as the one used for the TPB samples. A Perkin-Elmer Lambda~35~UV-Vis spectrometer with a Spectralon-coated IS (RSA-PE-20), equipped with a silicon photodiode (Hamamatsu S1227-66BQ operating in a photovoltaic mode), was used in the following configurations, see Fig.~\ref{fig:schematic}(a):
(1) T-mode, where the incident beam enters the IS after passing through the sample and (2) R-mode (0$^\circ$ or 8$^\circ$-wedged sample holder), where the beam reaches the sample after passing through the IS.
The standalone PEN film (a separate sample from the same batch earlier used in IS measurements), samples of either WLS on acrylic, and the bare acrylic substrate were measured in all three configurations.

Main systematic effects are directly probed by these configurations and varying sample surface finish. To first order, contribution of the reflected excitation beam is constrained by comparing 0$^\circ$~R- and T-mode measurements. Differences in angular distributions of fluorescence, as well as specularly reflected incident beam contribution are gauged by comparing 0$^\circ$ vs. 8$^\circ$~R-mode data.
Because there is no monochromator on the IS photodiode, in spectral regions where WLS effects are present the data are no longer interpreted as transmittance or reflectance, and instead become a product of an excitation-wavelength-dependent internal device calibration and the fluorescence intensity weighted by photodiode efficiency and Spectralon reflectance (both emission-wavelength-dependent), which can exceed 100\%.
Since spectra of TPB and PEN are similar, see Fig.~\ref{fig:spectra}(a), the remaining instrumental factors cancel out when taking the ratio, leaving only the relative performance of PEN and TPB. The residual uncertainty from spectral difference is estimated by integrating the product of emission spectra and photodiode efficiency to be $<$0.7\%, and neglected in the analysis.

The VUV setup with a 12.5$^\circ$ angle of incidence (as in \cite{Ouchi06}) and a fixed detection angle is composed of a deuterium lamp, a vacuum monochromator, and a light detector, see Fig.~\ref{fig:schematic}(b). The photon detector was either a custom-made photo multiplier tube (PMT) with an S20 cathode behind a MgF$_{2}$ window and an acrylic filter or a calibrated Ocean Optics QE65000 spectrometer. We used the spectrometer to measure the spectra of TPB and PEN shown in Fig.~\ref{fig:spectra}(a) and the PMT to measure the wavelength-integrated response of the samples. The reference sample had a 1.2$\pm$0.2~\micro m thick layer of TPB vacuum evaporated on glass. For geometrical consistency, the PEN sample was also clamped to a glass substrate, although not optically coupled (hence referred to as ``PEN (12.5$^\circ$)'').
The relative WLSE is determined based on the photon count rates of the PMT.
The dark rate is subtracted and the rates are corrected for pulse pile-up, dead-time of the data acquisition system and the wavelength-dependent response of the PMT and filter.
The error bars of the relative WLSE combine statistical uncertainty on the measured rates and systematic uncertainty, dominated by independently measured intensity fluctuations in the lamp. The error bars of the excitation wavelengths are standard deviations of the gaussian fits of the distribution of excitation light. Stray light made up at most 0.5\% of the signal. 

\subsection{Results}
Figure~\ref{fig:spectra}(b-d) shows the raw IS T- and R-mode spectra.
Relative WLSE of PEN to TPB is obtained by separately dividing the 0$^\circ$ R- and T-mode spectra below 250~nm by the TPB data, see Fig.~\ref{fig:extrapolation}(top). In doing so, the average value from all TPB samples is used, with the spread of values taken as the systematic uncertainty. The R-mode uncertainty is enlarged by the reflectance of the smooth substrate at 8$^\circ$ to account for any reflected incident excitation beam contaminating the fluorescence spectrum, while the T-mode uncertainty is inflated by the substrate transmittance. Additionally, for the case of standalone PEN (1) a 40\% solid angle correction is applied to T-mode data to account for the absence of the substrate, and (2) the R-mode systematic uncertainty is increased by the 0$^\circ$ and 8$^\circ$ data difference which is substantial, indicating sensitivity to the specularly reflected incident light.

Notably, WLSE for PEN+PMMA and for the T-mode measurement of standalone PEN are consistent. In addition to differences in angular distribution of reflected or emitted light, other reasons for the systematic shift of the R-mode standalone PEN results may include light losses due to scattering in the substrate.

The fixed angle VUV measurements can be systematically shifted compared to the IS measurement if angular distributions of emitted light differ between TPB and PEN; also, reflectance is wavelength dependent and for PEN can reach $\sim$20\%~\cite{Ouchi03}. An additional sample with a sanded front face was measured, where we expect (1) substantially reduced reflectance and (2) the emission angular distrubution much closer to TPB, although lack of the optical coupling with the substrate leads to front/back asymmetry. To account for these effects, the fixed angle data points are scaled by a factor providing the best match with weighted averages of the IS measurements at 210, 230 and 250~nm. The correction factors are 0.85(4) or 0.55(2) for the smooth and the sanded sample, respectively. After scaling both samples agree to 13\%, which is taken as a systematic uncertainty.

Next, relative WLSE spectra of PEN with respect to TPB are converted to absolute WLSE. This is done by scaling them by the absolute WLSE of TPB, using averaged values for 0.9 and 1.8~\micro\metre\ thick evaporated TPB coatings from Ref.~\cite{Benson}, see Fig.~\ref{fig:extrapolation}(bottom). 

\subsection{Projection to LAr conditions}
\begin{figure}[ht]
\centering
\includegraphics[trim={0 1.7cm -4.1mm 0},clip,width=0.784\linewidth]{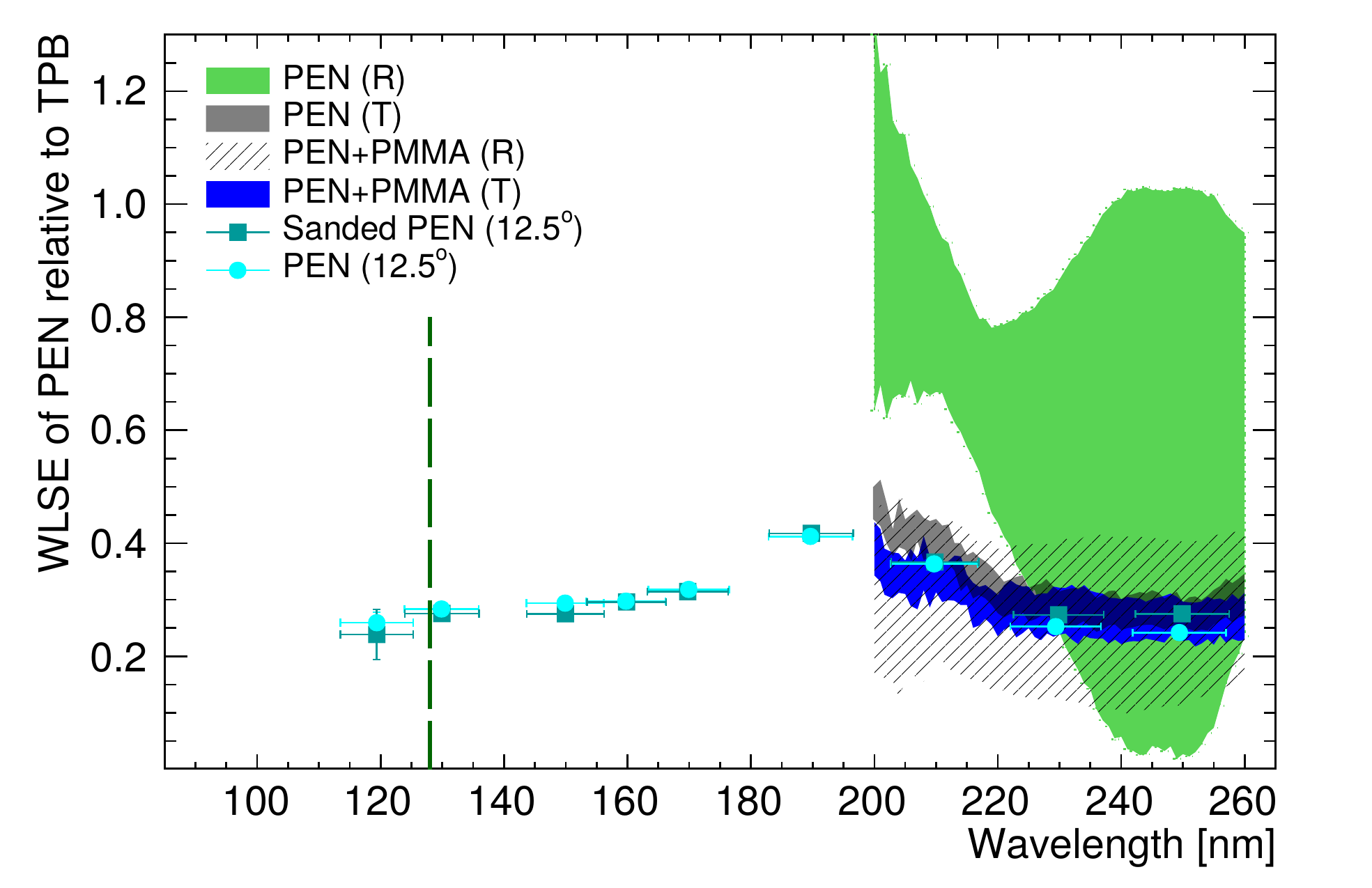}
\includegraphics[trim={0 0 -8mm 1.33cm},clip,width=0.78\linewidth]{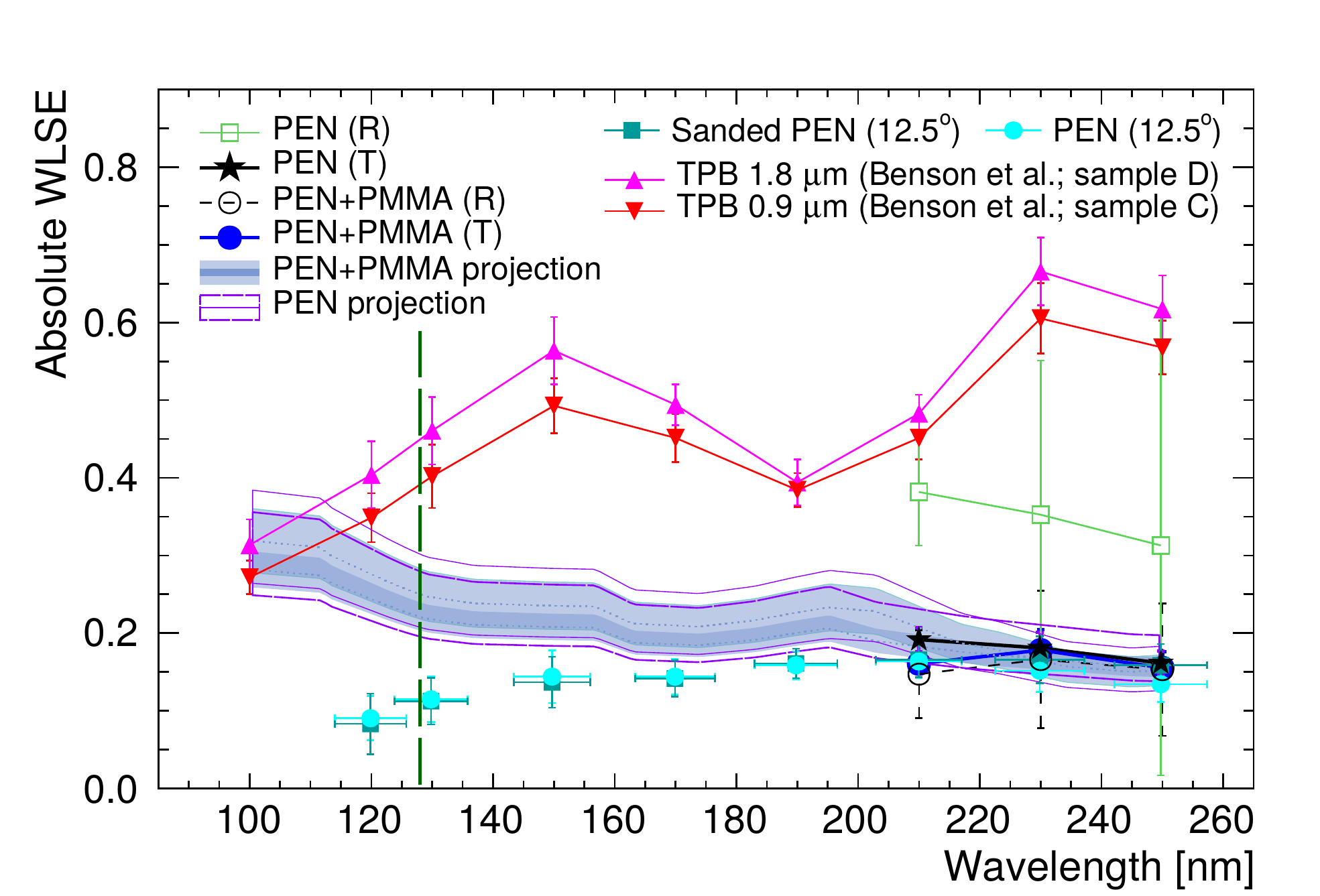}
\caption{Top: WLSE of standalone PEN and PEN on a substrate, relative to TPB on a substrate, measured in T- and R-mode IS configurations and with the VUV monochromator at 12.5$^\circ$ incidence (see legend). Data were taken at RT.
Bottom: comparison of measured and inferred absolute WLSE of PEN. Also shown are TPB data points used for the absolute calibration~\cite{Benson} (lines drawn to guide the eye). The vertical dashed line marks the 128~nm LAr scintillation wavelength. See text and legend for more details.
}
\label{fig:extrapolation}
\end{figure}

The WLSE of TPB can be now translated to LAr conditions based on the literature data: value at RT and 128~nm based on Fig.~\ref{fig:extrapolation}(bottom) and the 1.2 times increased value at 87~K based on Ref.~\cite[Fig.~9 therein]{Francini}.

Wavelength dependence of PEN WLSE is measured in this work with the normalization set by the IS measurements above 200~nm. Additionaly, data from Ref.~\cite[Fig.~5(b) therein]{Ouchi06} is used to project the IS measurements to 128~nm. The wavelength dependence of PEN photoluminescence intensity is the same at RT and 87~K in the entire 250--370~nm range explored by previous measurements, with the intensity approximately 1.7$\times$ higher at 87~K than at RT. The spectrum structure does not depend on the temperature or the excitation wavelength~\cite[Fig.~16 therein]{Laurent}\cite{Teyssedre}.

Interpolation of IS measurements down to 100~nm is performed by fitting the dependence of PEN fluorescence intensity on excitation wavelength~\cite{Ouchi06}, with the overall normalization factor kept as a free parameter. Two such fits are performed for the two extreme cases of PEN film draw axis orientation, for which a different dependence is expected. Fits are performed separately to 6 data points from PEN+PMMA and to 6 data points from the standalone PEN sample (R-mode data points do not affect the fit significantly because of large error bars).
The 95\% confidence bands of all four fits are extended down to 100~nm on Fig.~\ref{fig:extrapolation}(bottom), projecting PEN WLSE at 128~nm and enabling a direct comparison with TPB, as well as with the fixed angle VUV measurement.

Reasons for disagreement at lower wavelengths between PEN sample characterized in Ref.~\cite{Ouchi06} and VUV measurements of our Teonex\textsuperscript{\textregistered}~Q83 sample remain unclear. UV-induced degradation of PEN is one of possible mechanisms leading to such spread of results, in addition to varying level of impurities, cross-linking, or differences in polymeric chain parameters. Up to an order of magnitude deterioration of photoluminescence yield for excitation wavelengths in the 260--380~nm range as well as a shift of the peak emission wavelength towards higher wavelengths after 5~h of exposure to light from a 30~W broadband UVA lamp has been reported~\cite{Mary}. The effect is more pronounced at lower excitation wavelengths and strongly dependent on the length of exposure. Without detailed knowledge of the samples history it is impossible to evaluate the WLSE loss with respect to pristine PEN and correct for this effect; it is likely, however, that losses are generally higher at lower excitation wavelengths due to smaller penetration depth of VUV light and its higher sensitivity to the UV damage in the surface layer.

Table~\ref{tab:comparison} contains the resulting absolute WLSE values at 128~nm, conservatively extracted from Fig.~\ref{fig:extrapolation}(b) as the boundaries of the confidence bands (PEN) or error bar boundaries (TPB, considering both samples together).
Since the efficiency increase at low temperatures is usually related with reduced quenching from thermal non-radiative transitions~\cite{textbook}, which is independent of the excitation wavelength, the RT WLSE for PEN at 128~nm is projected to 87~K using the literature 1.7 enhancement factor~\cite{Laurent} (measured at 250~nm).
\begin{table}
\footnotesize
\centering
\begin{tabular}{l||cc|c}
    \multirow{2}{*}{Sample} & \multicolumn{2}{c|}{293~K} & 87--93~K \\
      & 250~nm & 128~nm & 128~nm \\
    
    \hline     {\footnotesize TPB+PMMA} & 0.59(7) & 0.43(7) & 0.52(10) \\\hline
    
{\footnotesize PEN+PMMA} & 0.15(2) & 0.24(4) & 0.40(7)  \\
{\small PEN} & 0.16(4) & 0.25(5) & 0.42(8)  \\
{\small PEN (glass)} & 0.15(3) & 0.12(3) & 0.20(6)  \\
   \hline
$\frac{\mathrm{PEN+PMMA}}{{\mathrm{TPB+PMMA}}}$ & 0.25(5)* & 0.56(13) & {\bf 0.77(20)} \\
        $\frac{\mathrm{PEN}}{\mathrm{TPB+PMMA}}$ & 0.27(8)* &  0.58(15) &  {\bf 0.80(23)}\\
        $\frac{\mathrm{PEN (glass)}}{{\mathrm{TPB+glass}}}$ & 0.26(4)* & 0.28(4)* & {\bf 0.38(7)} \\
\hline
$\frac{\mathrm{VM2000}}{\mathrm{TPB+TTX}}$ & 0.09 & &  0.317(16)\\
\end{tabular}
\caption{Absolute WLSE for TPB+PMMA~\cite{Benson}, PEN+PMMA, standalone PEN and PEN+glass, and relative efficiencies w.r.t. TPB+substrate. Values indicated with asterisks are directly measured in this work; the rest relies on literature based inference (see text for details). The last row shows experimental ratios of fluorescence yield of VM2000 and approx.~1~mg/cm\squared\ thick TPB coatings evaporated on Tetratex\textsuperscript{\textregistered} (TTX), with the 87~K value measured with an LAr detector from Ref.~\cite{Baudis}.
}
\label{tab:comparison}
\end{table}

With the above assumptions the projected WLSE of PEN in LAr in the worst case scenario, based on our own VUV measurement, is 38(7)\% of evaporated TPB, which already makes it competitive with the TPB-doped polymeric coatings considered for LAr neutrino detectors. The analysis leads to WLSE consistent with TPB when using wavelength dependence of a PEN sample from the literature, which, if confirmed by future measurements, would be important for dark matter detectors. Discrepancy between both methods indicates possible UV degradation effects or variability in WLSE between different PEN grades/batches, subject to future studies. Also, using a thinner PEN film than the one used in this test (125~\micro\metre) could bring further light yield enhancements.

To independently validate the results, we use published data directly comparing WLSE relative to TPB of PEN-containing VM2000 between 260~nm at RT and the LAr-based detector conditions, see Tab.~\ref{tab:comparison}. The absolute WLSE of ESR is expected to be significantly lower than that of pure PEN film due to UV light losses in layers of the non-scintillating polymer, but the relative comparison between both sets of conditions is meaningful. WLSE of VM2000 relative to TPB increases by a factor of 3.5(2) when going from 260~nm at RT to 128~nm at 87~K, which is consistent with factors of 3.0(1.1) and 3.0(1.2) inferred based on our IS results for PEN on acrylic and standalone PEN, respectively, projected to 128~nm using the data presented in Ref.~\cite{Ouchi06}. This direct empirical fact strongly supports our conclusions about the potential of PEN.

\section{Conclusions and outlook}
\label{S:2}
Using a compilation of literature data and a direct spectrometer measurement of WLSE of PEN relative to TPB, we find that the expected response of PEN to 128~nm excitation wavelength at 87~K is competitive with polymeric coatings doped with TPB, and may be close to that of evaporated TPB. 

While a direct experimental comparison of pristine samples in LAr is planned to remove the dependence on other published results, reduce the effects of UV-induced degradation and to identify the optimal material grade, the evidence collected in this paper provides a proof-of-principle, demonstrating the suitability of PEN for use as a WLS in LAr detectors.

Other essential characteristics, such as vacuum and cryogenic compatibility, stability, short fluorescence lifetime, and radiopurity have already been demonstrated in the literature, either for PEN or for ESR films.

The range of potential applications of PEN-based WLS is very broad. In particular, it appears feasible to replace TPB-coated ESR foils, which are under consideration for large LAr TPCs such as SBND~\cite{Szelc}, with a laminate of PEN and ESR\footnote{PEN could mitigate the TPB emanation in LAr~\cite{tpbdissolved} and find use in liquid Xe detectors, where TPB coatings dissolve~\cite{xetpb}.}.
In addition to long baseline neutrino and, if WLSE consistent with TPB is confirmed, dark matter projects, PEN could also find use in other LAr-based detectors: (1) CENNS-10~\cite{Tayloe} measuring coherent elastic neutrino-nucleus scattering, (2)  neutrinoless double $\beta$ decay searches employing LAr veto such as GERDA~\cite{gerda}\footnote{PEN laminated with inactive polymer (e.g. PET) could serve as an alternative  mitigation of \Nuc{Ar/K}{42} background~\cite{GerdaEnclosures}.} or the future LEGEND~\cite{legend}, (3) medical physics, (4) nuclear or homeland security.

Most importantly, PEN could directly replace TPB in the planned large ktonne-scale LAr experiments and dramatically simplify the WLS deposition step, which is currently one of the main technological bottlenecks.

\section*{Acknowledgements}
We thank Prof.~Mark Chen for access to the  spectrometer used for this work, and Teijin DuPont Films for the PEN samples. We are grateful to Dr.~Gilbert Teyssedre and Dr.~Dominique Mary for sharing details of their analysis. We also thank Dr.~Andrzej Szelc for valuable comments on the ma\-nu\-script. Support from Natural Sciences and Engineering Research Council of Canada and the McDonald Institute is gratefully acknowledged.

\end{document}